\documentclass[12pt]{iopart}

\usepackage{cite,epsfig,graphicx,amssymb}
\begin{document}

\title[Physics Opportunities at the LHC]{Opportunities for Heavy Ion Physics at the Large
Hadron Collider LHC}

\author{Urs Achim Wiedemann}

\address{Physics Department, CERN Theory group, CH-1211 Geneva 23}
\ead{Urs.Wiedemann@cern.ch}
\begin{abstract}
This talk discusses extrapolations to the LHC of several, apparently universal trends, seen in 
the data on relativistic nucleus-nucleus collisions up to RHIC energies. In the 
soft physics sector, such extrapolations to the LHC are typically at odds with LHC predictions
of the dynamical models, advocated to underlie multi-particle production up to RHIC energies.
I argue that due to this, LHC  is likely to be a discovery machine not only 
in the hard, but also in the soft physics sector. 
\end{abstract}


%
\section{Introduction}\label{sec:intro}

Heavy ion physics is an integral part of the baseline program at the CERN Large Hadron Collider
LHC~\cite{LHC-HI-exp}, which will start operation within the next year. In lead-lead collisions, 
the LHC will reach 
a center-of-mass energy of $\sqrt{s_{\rm NN}} = 5.5$ TeV, a factor 27 higher than the 
maximal energy explored at the Relativistic Heavy Ion Collider RHIC so far. This is an 
even larger increase in center of mass energy than the factor 10 in going from the CERN SPS 
to RHIC. It leads to a significant extension of the kinematic range in transverse 
momentum $p_T$ and in Bjorken-$x$, experimentally accessible for the study of hot and dense 
QCD matter. Rather than elaborating the often discussed novel questions, which we can address 
in the logarithmically wide {\it terra incognita} at high-$p_T$, this talk will focus on the most direct manifestation of QCD matter produced in heavy ion collisions: soft physics.

\section{LHC is a discovery machine for soft physics}
 
Soft multi-particle production has been studied extensively in the 
collisions of hadrons and nuclei, but despite insight from model-dependent approaches, it
is lacking a fundamental understanding.\footnote{This is so in p-p, as well as in A-A 
collisions. For instance, uncertainties 
in model extrapolations of the charged particle multiplicity per unit rapidity 
${\rm d}N_{\rm ch}/{\rm d}y$ over an order of magnitude in $\sqrt{s_{\rm NN}}$ are of similar
size in nucleus-nucleus and hadron-hadron collisions. Models successful
for hadron-hadron collisions up to the Tevatron energy ($\sqrt{s_{NN}} = 1.8$ TeV) vary by 
up to a factor 2 if extrapolated to p-p collisions at the LHC~\cite{Butterworth:2004is}. }  
In nucleus-nucleus collisions, soft multi-particle production shows imprints of collective
dynamics, such as elliptic flow, and of kinematic and hadrochemical 
equilibration~\cite{RHIC-white}. This
makes soft multi-particle production a testing ground for the central question of how 
collective phenomena, involving many degrees of freedom, can emerge from the 
fundamental laws of elementary particle physics.

\subsection{Example 1: multiplicity distributions}
\begin{figure}[h] 
\centerline{\includegraphics{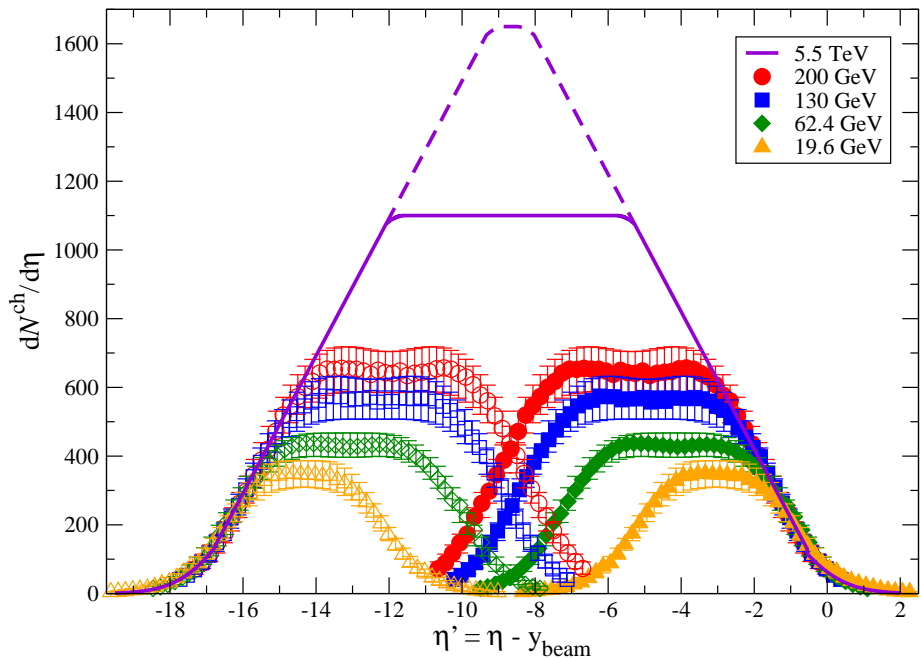}}
\caption{Pseudorapidity distribution of charged particle production in Au-Au collisions at
different center of mass energy. Data are plotted in the rest frame of one of the colliding nuclei
(full symbols), and mirrored at LHC mid-rapidity (open symbols). Agnostic extrapolations to the
LHC are based on assuming i)  that limiting fragmentation persists up to mid-rapidity(dashed line), 
or ii) that multiplicity distributions show limiting fragmentation and a self-similar trapezoidal shape 
(solid line). Data from~\cite{Back:2002wb,Back:2005hs}. Figure taken from~\cite{Nicolas}.
}\label{fig1}
\end{figure}

To illustrate the discovery potential of LHC for this class of questions, we contrast here
model predictions for heavy ion collisions at the LHC with extrapolations of generic 
trends. As repeatedly emphasized e.g. by Wit Busza~\cite{Busza:2004mc}, 
soft multi-particle production displays characteristic 
and apparently universal trends over many orders of magnitude in center of mass energy. 
For pseudo-rapidity distributions, these generic trends are i) extended longitudinal scaling 
and ii) the factorization of the $\sqrt{s_{NN}}$- and the centrality/A-dependence in
pseudo-rapidity distributions. Here, extended longitudinal scaling refers to the observation
that pseudo-rapidity distributions, plotted in the rest frame of one of the colliding hadrons, fall on a universal, energy-independent limiting curve in the projectile fragmentation region. The region within which this limiting fragmentation accounts for the data, increases with center of mass energy, see Fig.~\ref{fig1}.

%

Such generic trends can serve as a basis for agnostic extrapolations to the LHC. As seen in
Fig.~\ref{fig1}, requiring that both limiting fragmentation and the trapezoidal shape of the
pseudo-rapidity distribution persist at the LHC, one expects 
${\rm d}N_{\rm PbPb}^{\rm ch}/{\rm d}\eta \simeq 1100$ at $\eta = 0$. On the other hand, if one 
requires solely that the limiting fragmentation curve specifies the maximally allowed distribution 
at all rapidities, one concludes ${\rm d}N_{\rm PbPb}^{\rm ch}/{\rm d}\eta \simeq 1700$ at $\eta = 0$. 
In general, since the baseline of the trapezoid in Fig.~\ref{fig1} increases
$\propto \log\sqrt{s_{\rm NN}}$, limiting fragmentation implies that the multiplicity at
central rapidity increases at most logarithmically.

In marked contrast, a power-law increase of multiplicity distributions with  $\sqrt{s_{\rm NN}}$ is a 
generic consequence of perturbative particle production mechanisms, which may be expected to 
become relevant with increasing $\sqrt{s_{\rm NN}}$. Arguably, the main lesson learnt from the lower 
than predicted event multiplicities measured at RHIC~\cite{RHIC-white} is, that the 
power-law dependence of
(the simplest) perturbative multiplicity-enhancing mechanisms, such as minijet production,
is too strong to be reconciled with data. On the other hand, saturation models offer a 
fundamental reason for the very weak $\sqrt{s_{\rm NN}}$-dependence of event multiplicities, 
namely the taming of the perturbative rise due to density-dependent non-linear parton evolution.
They are based on the assumption that multiplicity distributions at ultra-relativistic energies are 
calculable within perturbation theory, since they are governed by a perturbatively high, 
$\sqrt{s_{\rm NN}}$- 
and $A$-dependent momentum (saturation) scale 
$Q_{\rm sat, A}^2 \propto \sqrt{s_{\rm NN}}^\lambda$. 
In saturation models, multiplicities at mid-rapidity rise essentially $\propto Q_{\rm sat, A}^2$ times 
transverse area. This also accounts naturally for the experimentally observed factorization
of $\sqrt{s_{\rm NN}}$- and centrality-dependence.
The exponent $\lambda$ is not a free fit parameter, but is taken to be 
constrained by data on $e-A$ collisions and by studies of non-linear small-$x$ evolution.
Depending on details of the modeling of this idea, one arrives at estimates 
between ${\rm d}N_{\rm PbPb}^{\rm ch}/{\rm d}\eta \simeq 1700$~\cite{Armesto:2004ud}
and ${\rm d}N_{\rm PbPb}^{\rm ch}/{\rm d}\eta \simeq 1800-2100$~\cite{Kharzeev:2004if}.
Similar values are also obtained by invoking other mechanisms to tame the
perturbative growth~\cite{Eskola:1999fc}. They are significantly higher than extrapolations of
the apparently universal trends shown in Fig.~\ref{fig1}.

To sum up this first argument: 
While the factor 30 increase in $\sqrt{s_{\rm NN}}$ from RHIC to LHC will not be
sufficient to discriminate a logarithmic increase from an arbitrarily tamed power-law increase,
it is sufficient to discriminate $\log \sqrt{s_{\rm NN}}$ from the factor
$\sqrt{s_{\rm NN}}^{\lambda}$, where $\lambda$ is constrained by our current
understanding of saturated QCD. So, we observe that the apparently universal trends 
seen in multiplicity distributions can be accounted for by saturation models up to RHIC 
energies. However, the logarithmic extrapolation of these trends to the LHC is not 
consistent with the power-law dependence of the dynamical models advocated to 
underly multi-particle production at RHIC. In this sense, LHC will be a discovery
machine for soft physics, starting from day 1 of its operation. Either, it will find characteristic
violations of the apparently universal trends, seen up to RHIC data - thus
providing qualitatively novel support for a specific microscopic collision dynamics. Or, 
LHC will confirm these generic trends - thereby prompting us to revisit the central 
dynamical ideas currently proposed for the tamed growth of event multiplicities up to
RHIC energies. In both cases, the day 1 measurement of multiplicity distributions at the LHC
is likely to have profound consequences for our understanding of the matter
produced in nucleus-nucleus collisions at the LHC {\it and} at RHIC.

\begin{figure}[t]
\centerline{\includegraphics[scale=1.2,angle=0]{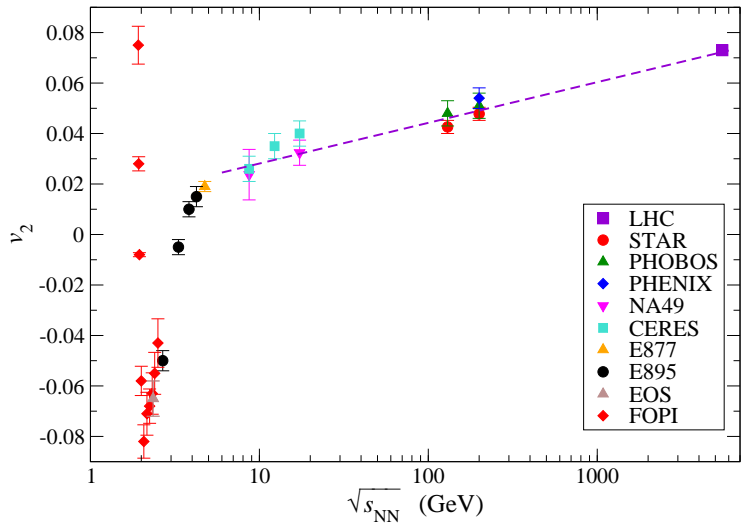}}
\caption{$\sqrt{s_{_{NN}}}$-excitation function of $v_2(y\!=\!0)$ in mid-central 
  collisions. 
  Data are taken from the compilation in Ref.~\cite{Alessandro:2006yt}.
  The point at LHC energy is obtained by agnostic linear extrapolation of the
  pseudo-rapidity dependence $v_2(\eta)$ from RHIC. Figure taken from~\cite{Nicolas}.}
\label{fig2}
\end{figure}

\subsection{Example 2: elliptic flow}
The above line of argument can be adopted to other characteristic features of
soft multi-particle production in heavy ion collisions. As a second illustration,
we mention here the azimuthal asymmetry of hadron production commonly
referred to as 'elliptic flow' $v_2$. The observable $v_2$ is constructed such
that it is an unambiguous signature of collective dynamics in the collision. At
lower energies, the $p_T$- integrated elliptic flow shows characteristic changes
of sign, indicative of significant changes in the collision dynamics.
At SPS energies and above, one finds that the pseudo-rapidity distribution
$v_2(\eta)$ of  $p_T$-integrated elliptic flow changes smoothly and shows 
extended longitudinal scaling~\cite{Back:2004zg}. The shape of
$v_2(\eta)$ is triangular, and extrapolation to the LHC yields $v_2(\eta=0)\simeq 0.075$. 

At RHIC energies, ideal fluid dynamic simulations of the collective dynamics arrive at
a fair description of elliptic flow at mid-rapidity~\cite{RHIC-white,Teaney:2001av,Kolb:2003dz}. 
They account for the absolute size of
$v_2$, its $p_T$-dependence up to $p_T \leq 1.5-2$ GeV, some aspects of the
centrality dependence and qualitative features of the particle-species dependence
~\cite{RHIC-white,Teaney:2001av,Kolb:2003dz}.
However, neither the approximately linear increase of $v_2(\eta=0)$ with
$\log\sqrt{s_{\rm NN}}$, nor the triangular shape of the pseudo-rapidity 
distribution of $v_2$ emerge as natural consequences of these dynamical models.
Even more restrictively, the energy dependence of detailed characteristics of elliptic
flow, such as PID-, $p_T$- and $A$- dependence, has not been employed fully to
test and refine fluid dynamic models, simply because one argues that an ideal
liquid is produced solely in sufficiently central collisions at the highest RHIC energies.
Thus, beyond the statement that RHIC may have seen the onset of ideal
fluid behavior, the main consequence of this claim is arguably the prediction that
this behavior will persist in heavy ion collisions above RHIC energies. 
Irrespective of whether LHC will finally be the confirmation or falsification
machine for this ideal fluid dynamic picture, it is clear that the factor 30 increase in
$\sqrt{s_{\rm NN}}$ turns LHC into the discovery machine, needed to adequately support such a
strong claim. LHC is well-positioned to provide critical tests for the ideal fluid paradigm.

The last argument may be questioned. It is true that ideal fluid simulations of heavy 
ion collisions at the LHC favor significantly lower elliptic flow values of 
$v_2(\eta=0) \simeq 0.055-0.06$
for $dN_{\rm ch}/d\eta \simeq 1100$, than the results of the extrapolation shown in Fig.~\ref{fig2}.
However, changes in the transverse spatial profile of the initial conditions, which are difficult
to constrain, have been found to affect the final $v_2$-signal significantly~\cite{Hirano:2005xf}.
One may suspect that such model-dependent uncertainties in the initial conditions
(or in implementing dissipative corrections and in simulating the freeze-out 
process~\cite{Baier:2006um}) are too
significant to turn LHC measurements into decisive tests which go qualitatively beyond what
has been achieved at RHIC. In my view, improving the accuracy of model simulations responds
only partially to such concerns. In addition, one should require that a 'good' dynamical
interpretation does not discard as mere numerical coincidences trends which persist over
wide kinematic ranges, but that it can account for them as natural consequences of the 
underlying dynamic picture. The fact that LHC extends for the first time by a factor $30$ 
the $\sqrt{s_{\rm NN}}$-range
within which ideal fluid dynamics applies (if it applies at RHIC), will give access to the
detailed study of such generic trends which must emerge from a valid dynamic explanation.
To what extent does the $\sqrt{s_{\rm NN}}$-dependence of $p_T$-integrated $v_2$ arise from
the increase of the rms transverse momentum $\langle \sqrt{p_T^2}\rangle$ with 
$\sqrt{s_{\rm NN}}$ in elementary nucleon-nucleon collisions (which is not directly 
invoked in ideal fluid dynamics)  rather than from an
increase of $v_2(p_T)$ at fixed $p_T$ with $\sqrt{s_{\rm NN}}$? How does $v_2$ at RHIC 
mid-rapidity differ from $v_2$ at an LHC rapidity which has the same $dN_{\rm ch}/d\eta$
as RHIC at $\eta = 0$? How does the breaking point in $v_2(p_T)$ and the PID
composition of $v_2$ change as a function of $\sqrt{s_{\rm NN}}$? The interest in these and
other systematic dependencies is clearly motivated by RHIC measurements. But
LHC will be the first machine to address these systematic dependencies in a 
kinematic regime in which ideal fluid dynamics is argued to apply. In this sense,
the physics opportunities at the LHC for critically testing the ideal fluid paradigm
go significantly beyond the obviously novel measurements, such as extending 
with the help of the azimuthal dependence of $D$-, $B$-meson and quarkonium spectra
our understanding of mass-ordering at small $p_T$ and of constituent quark counting rules
of $v_2$ at intermediate $p_T$.

\subsection{How do the properties of hot and/or dense QCD matter evolve?}

The heavy ion programs at RHIC and at the LHC are complementary for the understanding
of collective soft physics phenomena in ultra-relativistic heavy ion collisions. The arguments
presented above support this view by emphasizing that the $\sqrt{s_{\rm NN}}$-systematics provided
by LHC measurements is indispensable for providing compelling experimental support of
the main tentative physics conclusions reached at RHIC energies: the onset of saturation 
phenomena and the onset of ideal liquid behavior~\cite{RHIC-white}. More generally, 
with a combined 
analysis of data from RHIC and the LHC, the question will come into experimental reach
of how the fundamental properties of QCD matter, produced in ultra-relativistic heavy ion 
collisions, change with center of mass energy. A combined analysis of RHIC and LHC is
of particular interest, since no abrupt change of physics is expected to occur in between
the two collider energies. 

Here, we use the discussion of the $\sqrt{s_{\rm NN}}$- and $Y$-  dependence of the 
Cronin effect to illustrate the novel opportunities. The 
nuclear-modification factor for d-Au collisions, measured at RHIC, shows at intermediate
transverse momentum a characteristic Cronin-type enhancement at mid-rapidity, which is
typically attributed to initial state $p_T$-broadening. Towards forward rapidity, 
this enhancement disappears and one finds, compared to the yield in proton-proton 
collisions, a suppression which grows rapidly stronger with increasing rapidity 
$Y$~\cite{RHIC-white}. 
Models invoking non-linear small-x evolution of parton distributions arrive at a  
semi-quantitative description of this phenomenon. These models illustrate the fact that
extended longitudinal scaling, a precursor of QCD saturation at transverse momenta above 
the saturation scale $Q_s$, can account for a significant reduction of partonic yields at
intermediate $p_T$~\cite{Kharzeev:2002pc}. However, one may imagine other 
mechanisms at work in the
rapidity-dependence of $R_{dAu}(p_T,Y)$. In particular, any inelastic process which 
is enhanced due to multiple initial state scattering, has the potential to shift particle yield
from forward rapidity towards mid-rapidity. Since parton distributions are steeply falling
towards projectile rapidity, this may introduce a significant rapidity dependence of 
$R_{dAu}(p_T,Y)$. In the context of RHIC data, there has been some discussion on how
to disentangle such confounding factors from signals of non-linear QCD evolution, e.g.
by studying isospin effects and the particle species dependence of $R_{dAu}(p_T,Y)$.
However, the best discriminatory tool is arguably provided by the LHC: In models
of saturation physics, there is a one-to-one correspondence between effects of the 
rapidity dependence and the $\sqrt{s_{\rm NN}}$-dependence, simply because a parton distribution 
boosted to higher rapidity $Y$ is a distribution looked at in a process at higher $\sqrt{s_{\rm NN}}$. 
As a consequence, the saturation physics explanation of the $Y$-dependent disappearance
of the Cronin peak at RHIC also implies the replacement of this peak by a significant suppression
at LHC mid-rapidity. In contrast, the confounding mechanisms alluded to above are
expected to result in a small but visible Cronin-peak in $p$-$Pb$ collisions at LHC mid-rapidity.

In general, comparing the $Y$- and $\sqrt{s_{\rm NN}}$-dependence of measurements is a powerful
tool for identifying the manifestations of (non-linear) small-$x$ evolution, not only in
h-A but also in A-A collisions. One may go one step further and argue that it is of fundamental 
interest to
study how intrinsic properties of the produced matter, such as its dissipative characteristics
(e.g. shear viscosity, but not only shear viscosity), or its jet quenching parameter $\hat{q}$
depend on $\sqrt{s_{\rm NN}}$, and whether they depend on $\sqrt{s_{\rm NN}}$ solely 
via $dN_{\rm ch}/d\eta$.
This is so, since it is mainly the scale evolution, rather than predictions at a fixed scale, 
which provides tests of the fundamental QCD dynamics.~\footnote{We note that there has been
some progress recently on calculating {\it at fixed scale} and at strong coupling
fundamental quantities accessible in heavy ion collisions, such as $\hat{q}$ or the shear 
viscosity. For an overview of related results based on string theory techniques, see Ref.~\cite{Liu:2007ab}} It is an exciting thought that
the ability of comparing the $\sqrt{s_{\rm NN}}$- and $Y$-  dependence of measurements over
logarithmically wide ranges, well beyond providing a novel tool for discriminating 
different physics effects, may be the basis for a new line of fundamental scientific investigation 
into medium-dependent QCD evolution. 

\section{LHC is a discovery machine for hard physics}
On purpose, I have put the emphasis of this article on soft physics at the LHC. There is
a risk that these opportunities are overlooked or ranked second, simply because the 
{\it terra incognita} of hard probes, opening up at the LHC, is so striking. Novel 
opportunities for hard probes at the LHC have been emphasized
repeatedly~\cite{Accardi:2003gp} (for my own detailed view, see Ref.~\cite{Wiedemann:2005gm}). 
Here, I characterize them crudely
by recalling the following facts: First, high-$p_T$ hadron suppression at RHIC
persists unattenuated up to the highest transverse momenta (a factor 5 suppression in 
central collisions up to $p_T \sim 15-20$ GeV) tested so far~\cite{RHIC-white}. This makes it likely that
strong medium-effects will persist in heavy ion collisions at the LHC up to very high transverse
momenta in all aspects of hadron and multi-hadron production. Second, high-$p_T$ 
hadron production in the multi-10 to 100 GeV transverse momentum regime is a hard,
but abundant probe at the LHC. It is the size of the expected medium-dependent signal, which 
facilitates its unambiguous attribution to a specific dynamic attenuation mechanism despite 
the obvious experimental uncertainties  
in a high-multiplicity environment, and despite the well-known uncertainties of performing 
QCD calculations for this situation. And it is the abundance of the expected medium-dependent
signal, which will allow us to constrain details of the proposed dynamics against sufficiently
differential measurements. In short, the combination of a large signal and an abundant
yield is the basis for a detailed characterization of hot and dense QCD matter with hard
probes. 

Beyond this improved precision due to the wider kinematic range and higher yield, can 
we identify qualitatively novel aspects of QCD, which come into experimental reach at the LHC? 
Arguably, measurements of heavy ion collisions at the LHC will study for the first time
a logarithmically wide transverse momentum range at perturbatively high $p_T$. Since
internal jet structures (i.e. intra-jet multiplicity and energy distributions, as well as 
jet-like particle correlations) are known to be characterized by QCD evolution, this may
provide a unique possibility to test the medium-dependence of the QCD scale evolution. 
Also, the potential to discriminate from the underlying event entire jet structures
is significantly improved with increasing 
$\sqrt{s_{\rm NN}}$. 

There is at least one alternative line of thought which links the analysis of the microscopic
mechanisms underlying jet quenching to a novel fundamental question of hot and dense
QCD:  Jet quenching studies address the issue of how and to what extent equilibration
processes occur in heavy ion collisions. What could be further away from thermal
equilibrium initially than a more than 100 GeV parent parton? And how could we hope for
a clearer picture of how partons equilibrate, than by assessing in detail how this parent
parton approaches kinetic and hadrochemical equilibrium as a function of the external
parameters of the cauldron which we can regulate, such as in-medium 
pathlength or event multiplicity? Here, enhanced precision and enhanced kinematic range 
are likely to further novel connections between the phenomenology of heavy ion 
physics and the fundamental questions of hot and dense QCD. 

I am indebted to many friends and colleagues for arriving at some of the thoughts
expressed here. I would like to single out the collaboration with Nicolas Borghini 
on Ref.~\cite{Nicolas}, which has shaped part of  the present line of argument.


\end{document}